\documentstyle[11pt,appb,epsfig]{article} 

\newcommand{\be}{\begin{equation}}
\newcommand{\ee}{\end{equation}}
\newcommand{\ba}{\begin{eqnarray}}
\newcommand{\ea}{\end{eqnarray}}

\def\lsim{\mathrel{\rlap{\lower4pt\hbox{\hskip1pt$\sim$}}
    \raise1pt\hbox{$<$}}}     
\def\gsim{\mathrel{\rlap{\lower4pt\hbox{\hskip1pt$\sim$}}
    \raise1pt\hbox{$>$}}}     

\newcommand{\bce}{\begin{center}}
\newcommand{\ece}{\end{center}}

\def\lsim{\mathrel{\rlap{\lower4pt\hbox{\hskip1pt$\sim$}}
    \raise1pt\hbox{$<$}}}         
\def\gsim{\mathrel{\rlap{\lower4pt\hbox{\hskip1pt$\sim$}}
    \raise1pt\hbox{$>$}}}         

\begin{document}
\title{PROBING CHIRAL SYMMETRY RESTORATION WITH 
\protect\\HEAVY IONS
\thanks{Presented at the International Workshop
on 'The Structure of Mesons, Baryons, and Nuclei' on the occasion
of Josef Speth's 60th birthday}}
       
\author{JOCHEN WAMBACH
\address{Institut f\"ur Kernphysik\\
Technische Universit\"at Darmstadt\\
Schlo{\ss}gartenstr. 9\\ 
D-64289, Darmstadt, Germany\\}
\hspace{1cm}\\
RALF RAPP
\address{Department of Physics and Astronomy\\
State University of New York at Stony Brook\\
Stony Brook, NY 11794-3800, U.S.A.}}
\maketitle

\begin{abstract}

It is discussed how chiral symmetry restoration manifests itself through
mixing 
of vector and axial-vector correlators. The vector correlator is directly
accessible in relativistic heavy-ion collisions.  
Within models of the vector correlator its implications for low-mass dilepton 
spectra are reviewed.

\end{abstract}

\section{Introduction}

The goal of ultra-relativistic heavy-ion collisions is to create new states 
of hadronic matter that are believed to have existed 
until a few tens of microseconds after the 'big bang'. These efforts were
largely stimulated by lattice QCD predictions that hadronic matter
at high energy density undergoes a phase transition to a plasma of
deconfined quarks and gluons. At present the lattice calculations with
realistic
light quark masses yield a transition temperature 
$T_c=150\pm 20$ MeV~\cite{Laer}
at which there is a rapid increase of the thermodynamic quantities such
as energy- and entropy density. The transition is accompanied
by a restoration of chiral symmetry, signaled by a sharp decrease of
the chiral condensate, $\langle\bar q q\rangle_T$, near $T_c$. 

In the following, we shall concentrate on the chiral symmetry aspects
of the phase transition and potential signals in heavy-ion collisions. 
In the physical 
vacuum chiral symmetry is spontaneously broken. For the light meson spectrum 
this manifests itself in two ways: 
(i) the appearance of (nearly) massless Goldstone bosons
          (pion, kaon, eta) which interact weakly at low energies, and
(ii) the absence of parity dublets, {\it i.e.} the splitting of 
          scalar- and pseudoscalar, as well as vector and axial-vector 
          mesons.
For the following discussion property (ii) will be most important.
Given the vector and axial-vector currents ($N_f=2$),
\be
V_\mu^a=\bar q\gamma_\mu(\tau^a/2) q \ ,\qquad
A_\mu^a=\bar q\gamma_\mu\gamma_5(\tau^a/2) q \ , 
\ee 
the vacuum properties of the vector and axial-vector 
mesons are encoded in the corresponding correlators
\ba
\langle V_\mu^a(x)V^b_\nu(0)\rangle 
 & = & -{\delta^{ab}\over \pi}\int d^4q \ \theta(q^0) 
e^{iqx} \ {\rm Im}\Pi^V_{\mu\nu}(q)
\nonumber\\ 
\langle A_\mu^a(x)A^b_\nu(0) \rangle 
 & = & -{\delta^{ab}\over \pi}\int d^4q \ \theta(q^0) \ 
e^{iqx} \ \biggl ({\rm Im}\Pi^A_{\mu\nu}(q)-F_\pi^2 \ \delta(q^2) \ 
q_\mu q_\nu\biggr ) \ . 
\ea
Note the explicit contribution of the pion to the axial-vector correlator
through the pion weak decay constant, $F_\pi$.
Current conservation implies a four-dimensionally transverse tensor structure
\be
\Pi^{V,A}_{\mu\nu}(q)=\biggl (g_{\mu\nu}-{q_\mu q_\nu\over q^2}\biggr )
\Pi^{V,A}(s) \ , \qquad \Pi^{V,A}(s)={1\over 3}g^{\mu\nu}\Pi^{V,A}_{\mu\nu}(q)
 \ , 
\label{PiVA0}
\ee
where $s=q^2$.
The isovector part of the spectral function ${\rm Im}\Pi^V$ is observed 
from the hadronic part of
the $e^++e^-\to$ even $\pi$ which has a sharp resonance corresponding
to the $\rho$-meson (770 MeV). On the other hand, 
the isovector part of ${\rm Im}\Pi^A$ can be observed from the hadronic 
part of the decay $\tau\to$ odd $\pi$. This part has a broad $a_1$ peak
(1260 MeV).  The two spectral functions are clearly different which is
one of the experimental signatures that chiral symmetry is spontaneously
broken. The degree of symmetry breaking follows from the Weinberg sum rule
\cite{Wein}  
\be
{1\over \pi}\int {ds\over s}
\bigg ({\rm Im}\Pi^V(s)-{\rm Im}\Pi^A(s)\biggr )=F_\pi^2 \ . 
\ee
Physically it states that the difference in 
vector and axial-vector polarizabilities $\Pi^V_{\mu\mu}(0)-\Pi^A_{\mu\mu}(0)$
of the QCD vacuum is given by the order parameter of spontaneous symmetry 
breaking. A second sum rule can be derived
\be
{1\over \pi}\int ds
\biggl ({\rm Im}\Pi^V(s)-{\rm Im}\Pi^A(s)\biggr )=0 \ , 
\ee
which implies that the 'energy weighted' sum rules (EWSR) of the two 
spectral functions
are identical. This is a well-known consequence of the conservation of
vector and axial-vector currents in the chiral limit.

\section{How to Detect Symmetry Restoration?}
In the experimental study of chiral restoration in heavy-ion collisions 
the change in the quark condensate $\langle\bar qq\rangle_{\mu,T}$ is not 
directly measurable. However, it follows from chiral symmetry alone 
that, at the phase boundary,
the scalar and pseudo-scalar correlators as well as the vector and 
axial-vector correlators must become identical. 
In principle, the observation of parity 
mixing can thus serve as a unique signal for chiral restoration.
Since the vector and axial-vector correlators in the 
vacuum are largely saturated by narrow resonances, the objective is 
then to study the spectral changes of these 
'collective modes' as a function of $\mu$ and $T$.
 
The vector correlator is directly accessible in heavy-ion collisions since 
it couples to photons or dileptons both of which undergo negligible 
final-state interaction.  Concerning the spectral properties 
there are, in principle, 
two possibilities: (1) the vector modes could become 'soft' at the
phase boundary giving rise to 'dropping masses'.
This is the hypothesis of 'Brown-Rho scaling'~\cite{BR91} and would be a 
natural consequence of a direct relationship between masses and the chiral 
condensate, as found in the vacuum. It explains qualitatively the rapid 
increase in entropy density across
the phase boundary,  as is seen in lattice QCD~\cite{Karsch};  
(2) the vector mesons remain massive,  becoming degenerate 
with their axial partners. To elucidate this possibility further 
the properties of the in-medium correlators need to be discussed
in more detail.

\section{In-Medium Vector and Axial-Vector Correlators} 
In equilibrium, the finite-temperature (chemical potential) correlation 
functions are evaluated in the grand canonical ensemble,  
\be
\tilde \Pi^V_{\mu\nu}(q_0,\vec q)=i\delta^{ab}\int d^4x \ e^{iq\cdot x} 
\ \rho_i \ \langle{i}| V^a_\mu(x)V^b_\nu(0) |{i}\rangle \ ,  
\label{medcorr}
\ee
and a similar expression for $\tilde \Pi^A_{\mu\nu}(q_0,\vec q)$ 
($\rho_i$ is the usual density matrix). 
Specification of a matter rest frame implies that Lorentz invariance is broken. 
Thus the momentum-space correlators
$\tilde\Pi^{V,A}_{\mu\nu}$ will depend on energy and three-momentum 
separately and not on just their invariant $q$. Also, the number of Lorentz 
tensors is larger.
Introducing longitudinal and transverse projection tensors
$P_{\mu\nu}^{L,T}$ one has
\be
\tilde \Pi^{V,A}_{\mu\nu}(q_0,\vec q)=\tilde \Pi^{V,A}_LP^L_{\mu\nu}+
\tilde \Pi^{V,A}_T  P^T_{\mu\nu} \ , 
\ee
which in vacuum reduces to eq.~(\ref{PiVA0}) with 
$\Pi^{V,A}\equiv\Pi^{V,A}_L=\Pi^{V,A}_T$
The pion, being a massless Goldstone boson in the chiral limit, is special. 
It only contributes to the longitudinal axial correlator and,  without
loss of generality, can be subsumed in the axial-vector correlator. 

The in-medium extensions of the Weinberg-sum rules (for $\mu=0$) have 
been derived in~\cite{KaShu}. They read
\be
\int_0^\infty {dq_0 \ q_0\over q_0^2-\vec q^2}\biggl (
{\rm Im}\tilde \Pi_L^V(q_0,\vec q)
-{\rm Im}\tilde \Pi_L^A(q_0,\vec q)\biggr )=0
\label{medpol}
\ee
and
\be
\int_0^\infty dq_0 \ q_0 \ \biggl ( {\rm Im}\tilde \Pi_{L,T}^V(q_0,\vec q)
-{\rm Im}\tilde \Pi_{L,T}^A(q_0,\vec q)\biggr )=0
\label{medew}
\ee
and hold at each value of the three-momentum $\vec q$. Note that
the polarizability sum rule (\ref{medpol})
only involves the longitudinal part of the spectral functions
while the EWSR (\ref{medew}) holds for both parts separately.
As usual in many-body physics, sum rules put important constraints 
on models of the spectral functions and can be used to gain 
insight into the role of in-medium effects. This can be illustrated
most stringently in the case of vanishing baryo-chemical potential.
In the chiral limit, the pion is massless below the critical
temperature for chiral symmetry restoration and, in the low
temperature limit, the heat bath is dominated by pions. It can be 
proven that in this limit the masses of the vector and axial-vector
mesons do not change to order $T^2$~\cite{ElIo}. To this
order there are only changes in the couplings of the currents and the
finite-temperature correlators are described by $T$-dependent
mixing of the zero-temperature correlators
\ba
\tilde \Pi^V_{\mu\nu}(q_0,\vec q)&=&(1-\varepsilon) \ \Pi^V_{\mu\nu}(q)
+\varepsilon \ \Pi^A_{\mu\nu}(q)
\nonumber\\
\tilde \Pi^A_{\mu\nu}(q_0,\vec q)&=&(1-\varepsilon) \ \Pi^A_{\mu\nu}(q)
+\varepsilon \ \Pi^V_{\mu\nu}(q)  \  . 
\ea
To order $T^2$, $\varepsilon\equiv T^2/6 F_\pi^2$, and therefore the
temperature dependence of the pion decay constant is 
$F_\pi^2(T)=(1-\varepsilon)F_\pi^2$, consistent with the results from
chiral perturbation theory~\cite{GaLe}. As is easily verified,
the sum rules (\ref{medpol}) and (\ref{medew}) are fulfilled.
As the mixing becomes maximal, $\varepsilon=1/2$, chiral symmetry
is restored.  The in-medium sum rules can be used to gain
further insight into the behavior of the in-medium correlators.
Restricting the discussion to vanishing three-momentum for
simplicity and resonance saturation in the vacuum 
through the current-field identities
\be 
V^a_\mu={m^2_\rho\over g_\rho}\rho^a_\mu \ , \qquad 
A^a_\mu={m^2_{a_1}\over g_{a_1}}a^a_\mu+{\rm pion} \ , 
\ee 
we obtain
for the vector spectral density in the $\rho$-meson channel
\be
\pi{\rm Im}\tilde \Pi^\rho_L(q_0)={m_\rho^4\over g_\rho^2} \ {\rm Im}
{1\over q_0^2-m_\rho^2-\Sigma_\rho(q_0)} \ , 
\ee
where $\Sigma_\rho$ denotes the $\rho$-meson selfenergy at finite
temperature. A narrow width approximation, ${\rm Im}\Sigma_\rho
\ll {m_\rho^*}^2$, yields
\be
\pi{\rm Im}\tilde \Pi_L^\rho(q_0)={m_\rho^4\over g_\rho^2}
 \ \delta(q_0^2-m_\rho^2-{\rm Re}\Sigma_\rho(q_0)) \ . 
\ee
The pole mass is determined from ${m_\rho^*}^2=
m_\rho^2+{\rm Re}\Sigma_\rho(m_\rho^*)$ and the spectral density
can be written as 
\be
\pi{\rm Im}\tilde \Pi_L^\rho(q_0)=Z_\rho \ {m_\rho^4\over g_\rho^2} \ 
\delta ( q_0^2-{m_\rho^*}^2)
\ee
with a temperature-dependent residue.
Similarly, for the axial-vector spectral density, 
\be
\pi \ {\rm Im}\tilde \Pi^{a_1}_L(q_0)=Z_{a_1}{m_{a_1}^4\over g_{a_1}^2} \ 
\delta ( q_0^2-{m_{a_1}^*}^2)
+Z_\pi \ F_\pi^2 \ q_0^2 \ \delta(q_0^2) \ . 
\ee
Inserting these spectral densities into the sum rules
(\ref{medpol}) and (\ref{medew}) implies that $Z_\rho=Z_{a_1}$and  
\be
Z_\pi=2Z_\rho\biggl ({m_\rho^2\over {m_\rho^*}^2}-
{m_\rho^2\over {m_{a_1}^*}^2}\biggr) \ . 
\ee
From the fact that the vector- and axial-vector spectral functions have
to become identical at chiral restoration one expects in the narrow
width approximation that $m^*_{a_1}\to m_\rho^*$ as the phase
transition is approached. As a result the residue at the pion pole, 
$Z_\pi$, has
to vanish. Whether both ${m_\rho^*}$ and ${m_{a_1}^*}$ vanish at $T_c$
or remain finite cannot be decided from sum rule arguments alone.
To answer this question dynamical models have to be employed. 
 
\section{Hadronic Models for the Vector Correlator}
In modeling the in-medium vector and axial-vector correlators one
should realize that, at SpS energies, the phase space in the final state is 
dominated by mesons (mostly pions) with a meson/baryon 
ratio of 5-7~\cite{Stach}. 
An obvious starting point is therefore a pure pion gas. For the $\rho$-meson
which, in the Vector-Dominance Model (VDM), couples predominantly to two-pion
states this implies a modification of the pion loop through the heat bath
as well as contributions from direct $\rho -\pi$ scattering. 
The resulting thermal broadening has been calculated in various 
frameworks~\cite{pipiT} and found to be rather small. 
A model-independent approach has been put forward by \cite{Steele} which
relies on a chiral reduction formalism in the virial expansion. Being an
expansion 
in temperature, the in-medium vector correlator (\ref{medcorr}) 
can then be related to the
vacuum vector- and axial-vector correlators in much the same way as discussed
in the previous section. The vacuum spectral information is deduced from 
experiment. 

In spite of the scarcity of baryons in the hot hadron gas, they have a
significant impact on the spectral properties of vector mesons, largely
because of strong meson-baryon coupling. When baryons are involved
the mixing of vector- and axial-vector correlators is much more 
complicated than at $\mu=0$ and it has only started to be addressed 
recently~\cite{ChDeEr}. Putting these difficulties aside, several approaches 
have been put forward to determine the in-medium vector correlator. In
analogy 
to the virial expansion at finite temperature an obvious starting point
is a combined low $T$, small $\mu$ expansion of (\ref{medcorr}),
resulting in the leading-order corrections to the vacuum correlators
in both pion- and nucleon number density, $n_\pi$ and 
$n_N$~\cite{Steele,Kling}. 
To leading order 
in $n_N$ the nucleon Compton amplitude enters which is constrained from 
$\gamma N$ photoabsorption data and the nucleon polarizabilities. 
In~\cite{Kling} the nucleon Compton tensor is evaluated in the VDM combined 
with chiral $SU(3)$ dynamics, based on an effective meson-baryon Lagrangian.
The calculation reveals that the $\rho$-meson suffers substantial broadening, 
as density increases. The position of the 'pole mass' is hardly affected,
however.    

A second approach for the in-medium spectral properties of the 
 $\rho$-meson starts from the well-known observation that pion propagation 
in the nucleus is strongly modified. A wealth of elastic $\pi$-nucleus
scattering data has provided detailed understanding of the relevant
physical mechanisms~\cite{ErWe}. The dominant contributions originate
from $P$-wave $\pi N$ scattering through 
$N$-hole and $\Delta$-hole loops, giving rise to a momentum-softening of 
the pion dispersion
relation. In the VDM it is therefore natural to account
for this effect by replacing the vacuum two-pion loop with in-medium
pions~\cite{pipimed}. Gauge invariance is ensured by including 
appropriate vertex corrections. To lowest order in nucleon density 
this approach represents a pion cloud model for the nucleon Compton amplitude
which largely coincides with that of \cite{Steele} and \cite{Kling}. In 
addition to leading-order contributions in $n_N$ there naturally emerge higher
orders in density, most notably $n_N^2$ terms.
These correspond to two-nucleon processes of meson-exchange character
and $NN$- and $N\Delta$-bremsstrahlung contributions.
Besides the medium modifications caused by dressing the intermediate 
2-pion states, direct interactions of the $\rho$-meson with 
surrounding nucleons in the gas have to be considered. 
Indeed, there are several  
well-established resonances in the particle data table~\cite{PDG} 
which strongly couple to the $\rho N$ decay channel, e.g. the 
$N$(1720) and the $\Delta$(1905). 
This led to the suggestion~\cite{FrPi} to consider $P$-wave particle-hole 
excitations of the type $\rho N(1720)N^{-1}$ and 
$\rho\Delta(1905)N^{-1}$. In a more complete description other  
resonances with appreciable $\rho N$ widths have been 
included~\cite{PPLLM,RUBW}, most notably $S$-wave excitations into 
$N(1520)$, $\Delta(1700)$, etc.. 
The most simple version of the VDM tends to overestimate the 
$B^*\to N\gamma$ branching fractions when using the hadronic coupling 
constants deduced from the $B^*\to N\rho$ partial widths. This is  
corrected for by employing an improved version of the VDM~\cite{KLZ},
which allows to adjust the $B^*N\gamma$ coupling independently.

\begin{figure}
\psfig{figure=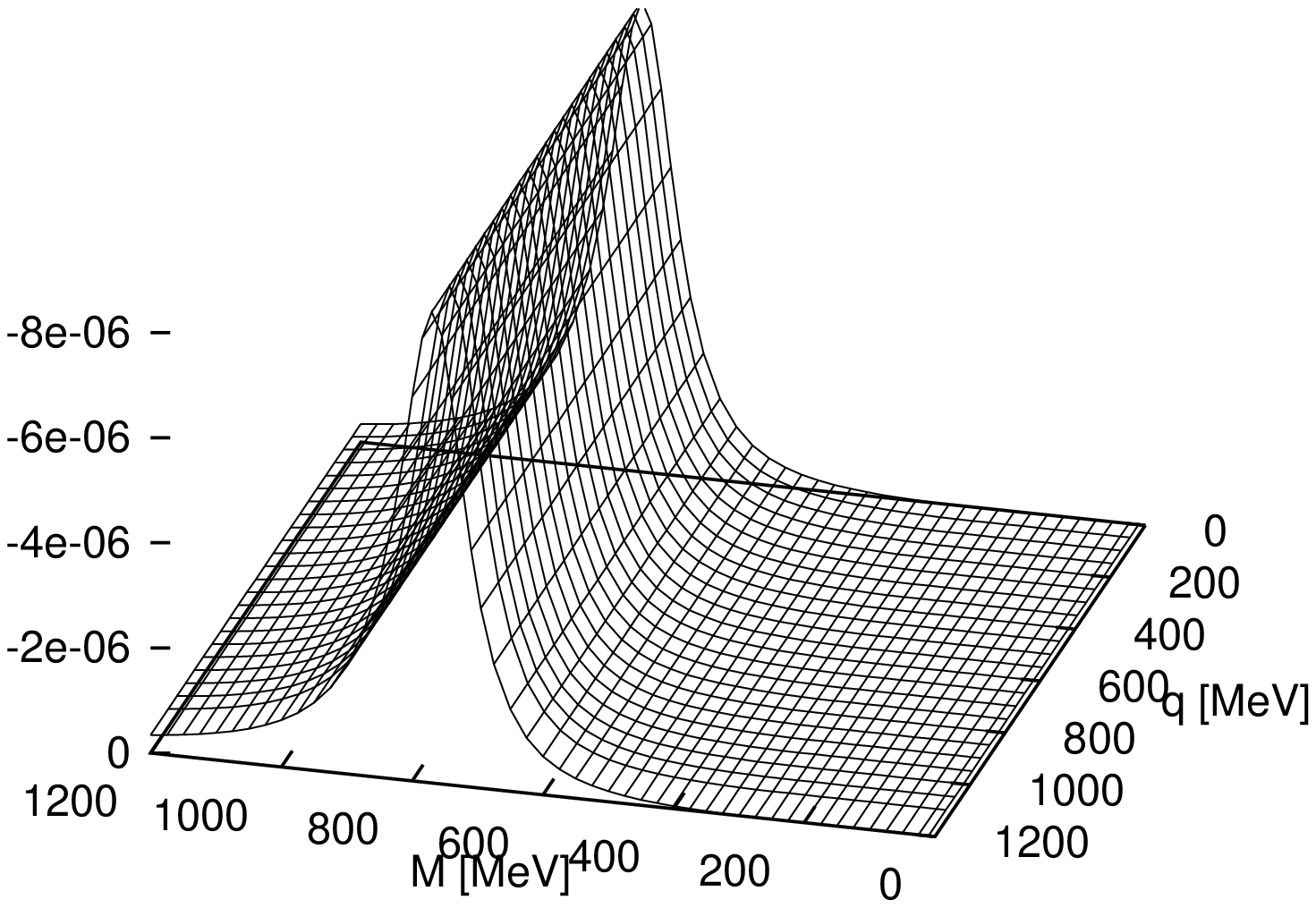,height=6.25cm,width=5cm,angle=-90}
\psfig{figure=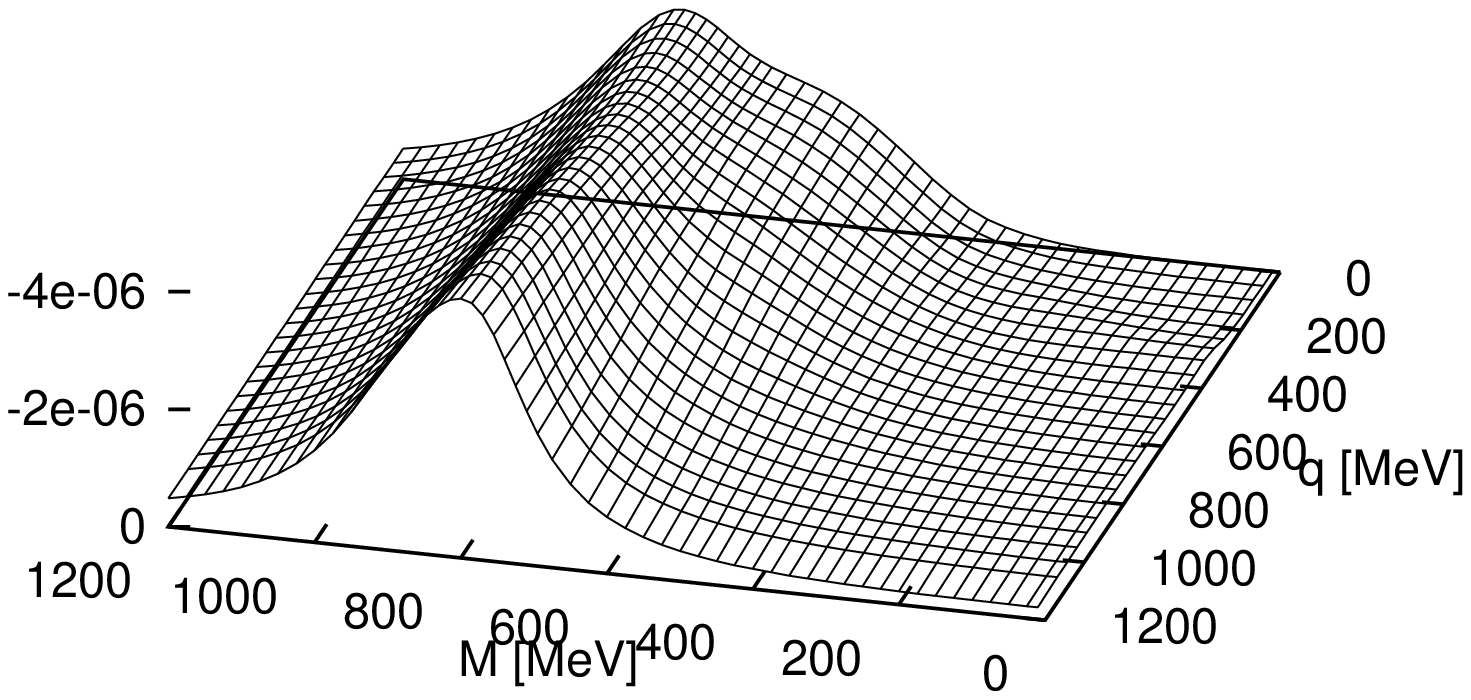,height=6.25cm,width=5cm,angle=-90}
\caption{$\rho$-meson spectral function versus invariant mass $M$ and 
3-momentum $q$ in vacuum (left panel) and in normal
nuclear matter (right panel).
\label{fig:matten}}
\end{figure}

Combining the effects of pionic modifications and resonant $\rho N$
scattering, the resulting $\rho$-meson spectral functions 
${\rm Im}\tilde \Pi^\rho=1/3({\rm Im}\tilde \Pi_L^\rho+2{\rm Im}\tilde \Pi_T^\rho)$ 
are displayed in Fig.~\ref{fig:matten}.
One observes significant broadening especially at small $\vec q$ with
a low-mass shoulder which originates from intermediate $\Delta N^{-1}\pi$
states and resonant $N(1520)N^{-1}$ excitations.
As $\vec q$ increases the shoulder moves towards the $M=0$ line which
can be understood from simple kinematics.  

\begin{figure}
\psfig{figure=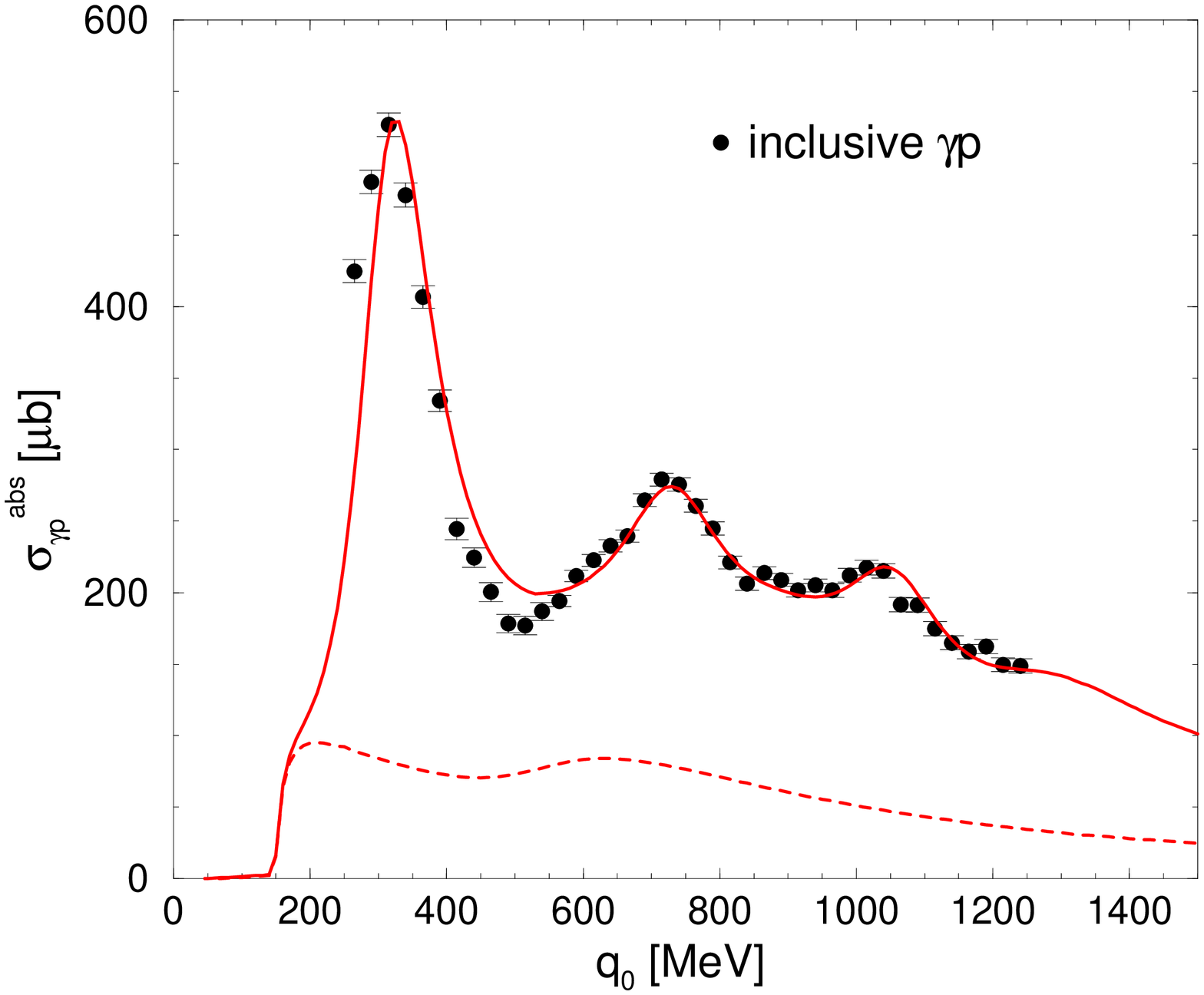,height=6.25cm,width=5cm,angle=-90}
\psfig{figure=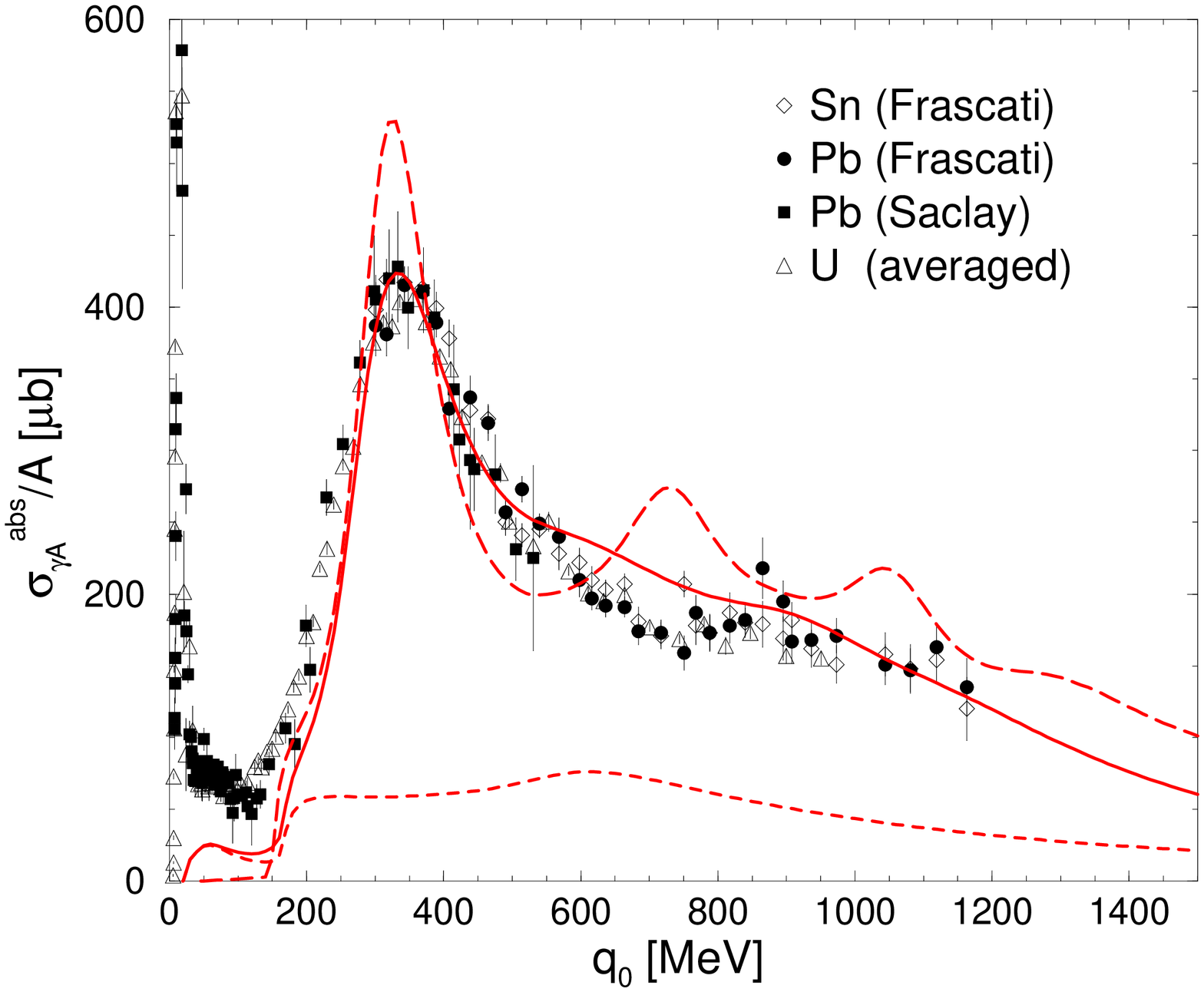,height=6.25cm,width=5cm,angle=-90}
\caption{The photo absorption spectrum on the proton~\protect\cite{photop}
(left panel) and on nuclei~\protect\cite{photoA} (right panel). The solid  
lines are the full results in the low-density limit (left panel) and for  
$\rho_N=0.8\rho_0$ (right panel), and short-dashed lines indicate the 
non-resonant background contributions; the results are taken 
from~\protect\cite{morio}.}
\label{fig:photo}
\end{figure}

An important constraint for models of in-medium $\rho$-meson propagation
is photoabsorption on nucleons and nuclei, for which a wealth of  
data exist over a wide range of energies. 
Real photons correspond to the 
$M=0$ line in the right panel of Fig.~\ref{fig:matten}. Within the
model discussed above, the total photoabsorption cross section per
nucleon can be calculated straightforwardly~\cite{RUBW}.
Taking the low-density limit, $n_N\to 0$, only terms
linear in density contribute, representing the absorption process
on a single nucleon. 
Adjusting the model parameters to optimally reproduce the $\gamma p$
data yields results displayed in the left panel of 
Fig.~\ref{fig:photo}. Photoabsorption on nuclei can be reproduced with
similar quality (right panel of Fig.~2).
It is noteworthy that, in the nucleus, strength below $m_\pi$ is obtained
which originates from two-nucleon processes via meson-exchange currents and is
nothing but the 'quasi-deuteron tail' of the giant dipole resonance.    
  
Another model constraint can be obtained from the analysis of 
$\pi N\to \rho N$ production~\cite{Fri97} which is in fact dominated 
by pion cloud contributions (rather than $B^*$ resonances). This imposes
stringent constraints on the hadronic form factor at the $\pi NN$ vertex
(requiring $\Lambda_{\pi NN}\simeq 300-400$~MeV), which are  
included in the results shown in Figs.~1-3~\cite{morio,RBRW}.  

\section{Comparison with Dilepton Data}

Including the above mentioned hadronic model constraints gives
confidence in extrapolations to the time-like region of dilepton production.
For $\pi\pi$-annihilation/$\rho$-meson decays the in-medium dilepton rate is 
obtained as
\be
{dN_{\pi^+\pi^-\to l^+l^-}\over d^4xd^4q} =
-\frac{\alpha^2}{3\pi^3} \
\frac{f^\rho(q_0;T)}{M^2} \ g^{\mu\nu} \ {\rm Im}\tilde\Pi^\rho_{\mu\nu}
(q_0,\vec q)\, .
\label{rate}
\end{equation}
In the model of \cite{morio,RCW} most of the important processes
discussed above have been included such that the $\rho$-meson propagator 
contains aside from the nucleonic contributions also 
some parts from the pion/kaon gas.

\begin{figure}
\psfig{figure=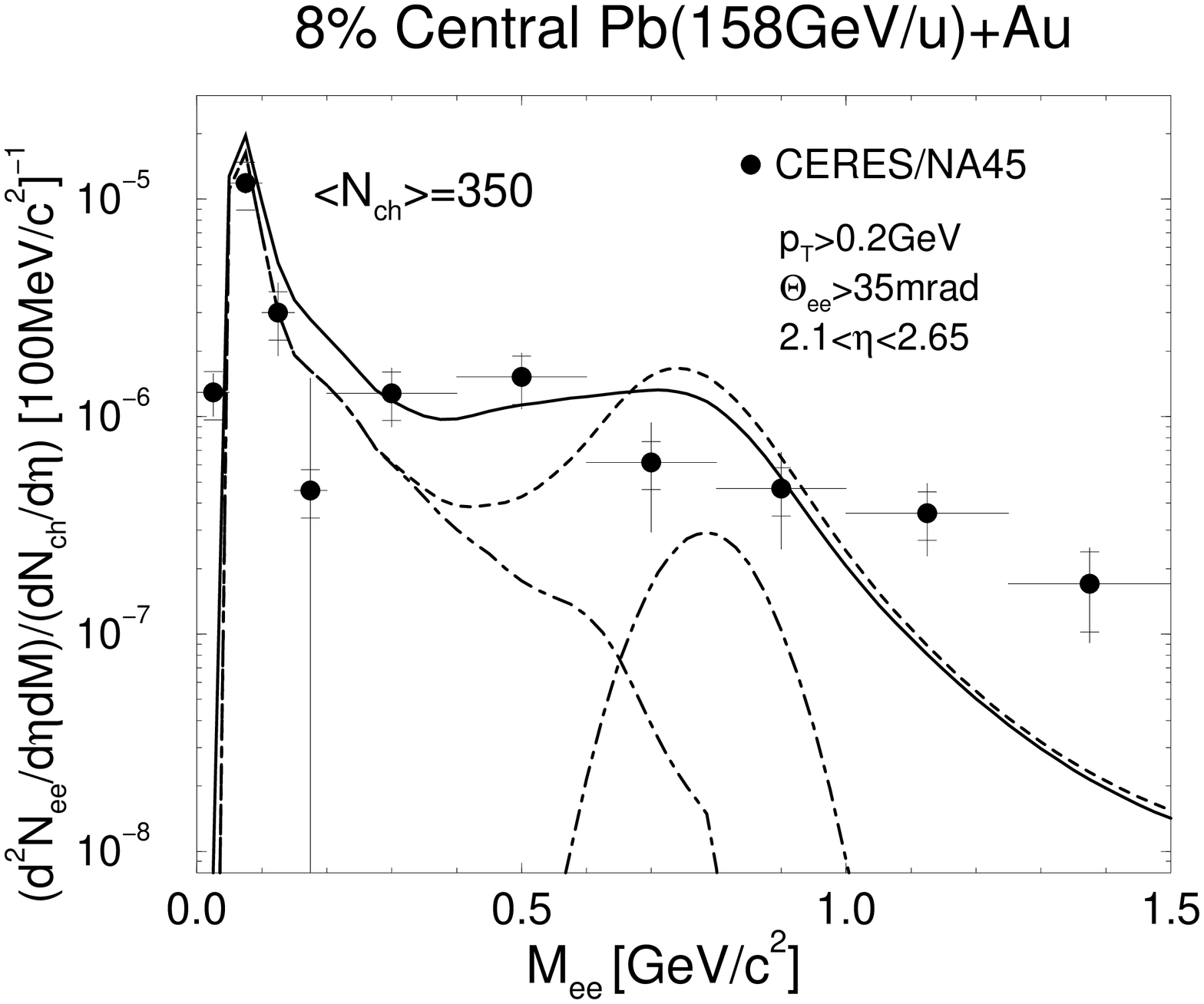,height=6.1cm,width=5cm,angle=-90}
\psfig{figure=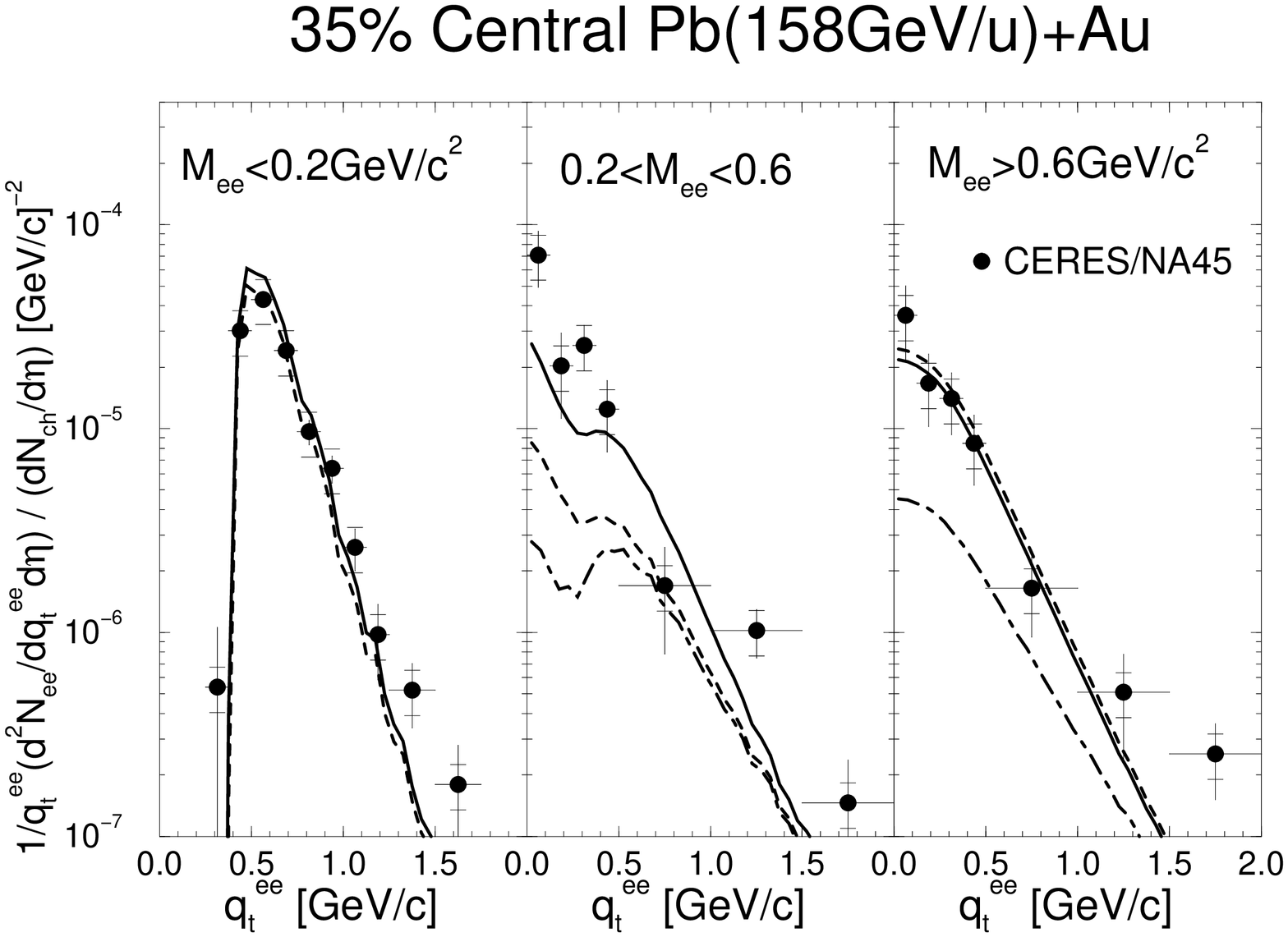,height=6.1cm,width=5cm,angle=-90}
\caption{Dilepton invariant mass (left panel) and
transverse momentum (right panel)
spectra in Pb+Au collisions at CERN-SpS
energies~\protect\cite{cerespt}. The dashed-dotted lines arise   
from hadron decays after freezeout~\protect\cite{cerespt,LKB}; 
adding the contribution from $\pi^+\pi^-$ annihilation in the hadronic 
fireball, one
obtains the dashed lines (when using the free $\rho$ propagator)
or the full lines (when using the in-medium $\rho$ 
propagator); the results are taken from \protect\cite{morio}.
\label{dlspec}}
\end{figure}

To compare the theoretical rates with data, the space-time history
of the heavy-ion collision has to be specified. There are
several possibilities for modeling the collision. Within  
a simple 'fireball' model~\cite{CRW,RCW} initial 
conditions in temperature and hadron abundances 
are taken from transport model calculations;  assuming local thermal
equilibrium as well as chemical equilibrium, the space-time history
is then determined by a simple cooling curve, $T(t)$, from some
initial time, $t_i$, up to the 'freeze-out time', $t_f$.  For SpS
energies such cooling curves are available from transport theory~\cite{LKB} 
and can be easily parameterized. The time evolution of the hadron abundances 
is determined by chemical equilibrium and agrees well with the 
transport model results. The observed spectrum is then obtained by 
integrating the 'local rate' (\ref{rate}) in time, accounting for the
detector acceptance in addition, cp. Fig.~\ref{dlspec}. 
When including the in-medium
effects due to hadronic rescattering as discussed above, reasonable
agreement is obtained (full curves) with both the invariant mass (left panel)
and transverse momentum spectra (right panel). Note that the major part
of the enhancement for 0.2~GeV$<$$M_{ee}$$<$0.6~GeV is correctly ascribed
to rather small pair momenta $q_t$$\leq$0.7~GeV. This might in fact resolve
the question why the $\mu^+\mu^-$ spectra measured by the NA50 collaboration
show a much less pronounced excess at low $M_{\mu\mu}$: their transverse
momentum cut of $p_t$$>$1~GeV will eliminate most of
enhancement generated within the model.

The fireball model is rather crude and does not incorporate detailed 
flow dynamics. A more sophisticated description is provided
by hydrodynamical simulations. In this case the 'local rates' (\ref{rate})
refer to a given fluid cell and can be directly implemented. So far, however,
no results are available.

A third way is to supply transport calculations with rates 
obtained from in-medium vector-meson propagation. For the $\rho$-meson
these have been implemented in simulations using the $HSD$ model for
the transport~\cite{CBRW}. The results are very similar to the ones of
Fig.~3 obtained in the fireball model.

\newpage

\section{Summary}

Recent theoretical efforts in understanding the nature of chiral 
symmetry restoration at
finite temperature and baryochemical potential and their implications for
low-mass dilepton production in URHIC's have been discussed. Special
emphasis has been put on the in-medium Weinberg sum rules and their
implications for the mixing of vector and axial-vector correlators. 
A unique signal for chiral symmetry restoration in URHIC's
would be the observation of such a mixing.
It strictly follows from chiral symmetry but is
difficult to detect. Only the vector correlator is accessible
via electromagnetic probes.

Meanwhile, the development of hadronic models for in-medium properties of 
vector mesons has advanced to a more quantitative level, in particular 
owing to phenomenological constraints inferred e.g. from photoabsorption 
data. Thus, trustworthy calculations for dilepton production in hot/dense
matter can be performed. While some approaches
inherently involve constraints from chiral symmetry~\cite{Kling,Steele}, 
others lack an obvious connection to chiral symmetry emphasizing, however,
input from hadronic phenomenology~\cite{CRW,RCW,RUBW}.  When supplemented 
by models for the space-time history of the heavy-ion collision dynamics, 
reasonable theoretical account for the experimentally observed low-mass 
dilepton enhancement seems to emerge~\cite{RCW,CBRW,morio}. 
Further calculations addressing more exclusive
observables as e.g. the recently measured transverse momentum 
spectra~\cite{cerespt}, where the major part of the low-mass enhancement 
has been identified at low pair-$p_t$, also seem to be in line~\cite{morio} 
with the data. 

\section*{Acknowledgments}
This work was supported in part by 
the A.-v.-Humboldt foundation (within a Feodor-Lynen fellowship),
the NSF  under grant no. NSF-PHY-94-21309 and
the US-DOE under grant no. DE-FG02-88ER40388.

\end{document}